\documentclass[aps,prb,twocolumn]{revtex4-2}

\usepackage{amssymb}
\usepackage{graphicx}
\usepackage{epsf,epsfig}
\usepackage{bm}
\usepackage{bbm}
\usepackage{amsfonts}
\usepackage{amsmath}
\usepackage{graphics}
\usepackage[normalem]{ulem}
\usepackage{xcolor}
\usepackage{hyperref}

\newcommand{\be}{\begin{eqnarray}}
\newcommand{\ee}{\end{eqnarray}}
\newcommand{\ba}{\begin{array}}
\newcommand{\ea}{\end{array}}

\newcommand{\tr}{\mbox{tr}}
\newcommand{\Tr}{\mbox{Tr}}

\newcommand{\eps}{\varepsilon}

\newcommand{\bfr}{{\bf r}}

\newcommand{\bfq}{{\bf q}}
\newcommand{\bfk}{{\bf k}}

\begin{document}

\title{Interaction corrections to the thermopower of the disordered two-dimensional electron gas}

\author{Zahidul Islam Jitu}
\author{Georg Schwiete}

\affiliation{Department of Physics and Astronomy, The University of Alabama, Tuscaloosa, Alabama 35487, USA}

\begin{abstract}
At low temperatures, the transport coefficients in the disordered electron gas acquire quantum corrections as a result of the complex interplay of disorder and interactions. The interaction corrections to the electric conductivity have their origin in virtual processes with typical electronic energies far exceeding the temperature. Here, we study interaction corrections $\delta S$ to the thermopower $S$ of the two-dimensional disordered electron gas with long-range Coulomb interactions. We show that while both real and virtual processes contribute to these corrections, the real processes are dominant and lead to a logarithmic temperature dependence of $\delta S/S$ with $\delta S/S<0$. 
\end{abstract}

\maketitle

{\it Introduction.---} 
The low-temperature transport properties of disordered electron liquids have fascinated researchers for decades \cite{AltshulerLee80,Finkelstein83,Altshuler85}, for a recent review see Ref.~\cite{Finkelstein23}. Transport anomalies in quantum critical metals have recently triggered a renewed interest in the intricate interplay of disorder and electron-electron interactions characteristic for these systems \cite{Nosov20,Wu22,Guo22,Patel23}. Despite the long research history on disordered electronic systems, theoretical studies of interaction corrections to thermoelectric transport are rather rare \cite{Hsu89,Fabrizio91}. Yet, compared to electric and thermal transport, the thermoelectric transport coefficient offers complementary information as it probes different aspects of the electron dynamics. In this manuscript, we investigate thermoelectric transport at low temperatures in the context of the two-dimensional disordered electron gas with long-range Coulomb interactions. 

The charge current ${\bf j}$ and the heat current ${\bf j}_k$ flowing in a system of electrons in response to an electric field ${\bf E}$ or a temperature gradient $\nabla T$ are characterized by a matrix of transport coefficients \cite{Ziman01,a_RemarkThermal} as
\begin{align}
\left(\ba{cc} {\bf j}\\ {\bf j}_k\ea\right)=\left(\ba{cc} \sigma &\alpha \\\alpha T&\kappa\ea\right) \left(\ba{cc} {\bf E}\\ -\nabla T\ea\right).\label{eq:matrix}
\end{align}
In a conventional Fermi liquid, the electric conductivity $\sigma$ and the thermal conductivity $\kappa$ fulfill the Wiedemann-Franz law \cite{Wiedemann1853} $\kappa=\mathcal{L}\sigma T$ at low temperatures $T$, where $\mathcal{L}=\pi^2/3e^2$ is the Lorenz number, and $e$ is the electron charge \cite{Units}. The off-diagonal matrix elements in Eq.~\eqref{eq:matrix} differ only by a factor of temperature, a manifestation of the Onsager relation known from non-equilibrium statistical mechanics \cite{Onsager31,Onsager31a,vanVliet08}. The coefficient $\alpha$ governs thermoelectric response. In experiment, it can be found from measurements of the Seebeck coefficient $S=\alpha/\sigma$, which is also known as thermopower.

Due to the combined effect of disorder and interactions, the electric conductivity of the disordered electron gas acquires quantum corrections at low temperatures \cite{AltshulerLee80,Finkelstein83,Altshuler85} . These interaction corrections become particularly strong when the temperature is small compared to the impurity scattering rate $T\ll 1/\tau$. The inequality $T\ll 1/\tau$ characterizes the diffusive transport regime, which we will focus on in this manuscript. In two dimensions ($2d$), the correction to the electric conductivity caused by the Coulomb interaction is logarithmic and takes the universal form \cite{AltshulerLee80}
\begin{align}
\delta \sigma=-\frac{e^2}{2\pi^2}\log\frac{1}{T\tau}.\label{eq:dsigma}
\end{align}
This correction originates from virtual processes with electronic energies in the interval $(T,1/\tau)$, the energy interval that also underlies the renormalization group (RG) analysis of the disordered electron gas \cite{Finkelstein83,Castellani84}.

The thermal conductivity $\kappa$ acquires interaction corrections from the RG energy interval 
as well \cite{Castellani87}, in accordance with the Wiedemann-Franz law. In addition, logarithmic corrections to $\kappa$ arise from real processes, specifically from the sub-thermal energy interval $(T^2/D\kappa_s^2, T)$, where $\kappa_s$ is the inverse screening radius and $D$ is the diffusion coefficient \cite{Livanov91,Raimondi04,Niven05,Catelani05,Michaeli09,Schwiete16a,Schwiete16b}. These corrections 
violate the Wiedemann-Franz law by increasing the thermal conductivity disproportionately 
\begin{align} 
\frac{\delta\kappa}{\kappa}=\frac{\delta \sigma}{\sigma}+\frac{1}{2}I^h,\label{eq:dkappa}
\end{align}
where 
$I^h=\rho \log (D\kappa_s^2/T)$, $\rho=1/(4\pi^2 \nu_0 D)$ is the dimensionless resistance, $\nu_0$ is the density of states per spin direction, and the classical Drude result $\sigma=2e^2 \nu_0 D$ implies $\delta \sigma/\sigma=-\rho \log 1/T\tau$. It is worth mentioning that in a model system where particles interact via short-range Fermi liquid-type amplitudes, logarithmic corrections arise from the RG energy interval only, and the Wiedemann-Franz law holds \cite{Castellani87,Schwiete14a}.

{\it Results.---} In this manuscript, we develop a theory of interaction corrections to the thermopower of the $2d$ disordered electron gas with long-range Coulomb interactions \cite{b_RemarkAmplitudes}. We show that the thermopower $S$ acquires logarithmic corrections from the RG energy interval as well as from subthermal energies. We find the following result
\begin{align}
\frac{\delta S}{S}=-\frac{1}{4}\frac{\delta \sigma}{\sigma}-I^h.\label{eq:alphaS}
\end{align}
The logarithmic corrections represented by $\delta \sigma/\sigma$ originate from the RG energy interval, and those entering via $I^h$ from subthermal energies. In view of the localizing character of the interaction corrections to the electric conductivity, $\delta \sigma/\sigma<0$, and due to $I^h>0$, the two corrections to the thermopower partially cancel. Remarkably, the correction from the subthermal energy interval dominates and thereby determines the overall sign of $\delta S$. Thus, real processes carry a larger weight for the corrections to the thermopower than virtual processes. This is in sharp contrast to the electric conductivity, for which the corrections are caused by virtual processes only, while real processes do not contribute at all. Overall, the result for the interaction corrections to the thermopower can be written as 
\begin{align}
\frac{\delta S}{S}=-\frac{3}{4}\rho\log\frac{\bar{E}}{T}<0,
\end{align}
where $\bar{E}=(D\kappa_s^2)^{4/3}\tau^{1/3}$ serves as an effective energy scale.  Corrections to the thermopower from the RG energy interval have been addressed in the seminal work of Ref.~\cite{Fabrizio91}. However, our results differ from those obtained in this reference, as well as the earlier Ref.~\cite{Hsu89} (see remark~\cite{c_RemarkHsuFabrizio1}). In particular, the role of the sub-thermal energy interval was not recognized in these previous studies. Here, we find that corrections from the sub-thermal energy interval change the overall sign of the correction to the thermopower compared to the prediction of Ref.~\cite{Fabrizio91}.

{\it The role of particle-hole asymmetry.---}
There is a crucial difference between the diagonal elements of the matrix of transport coefficients in Eq.~\eqref{eq:matrix}, and the off-diagonal ones describing thermoelectricity. Thermoelectric transport is a sensitive probe of particle-hole symmetry in electronic systems \cite{Ziman01}. Since the response of particles to a non-uniform temperature is independent of the sign of their charge, the resulting thermopower vanishes in particle-hole symmetric systems. This implies that the energy-dependence of parameters like the density of states, electron velocity and the momentum relaxation time, which requires particle-hole asymmetry, becomes crucial. For instance, in metals these parameters are almost constant as a function of energy up to corrections of the order of $T/\mu$, where $\mu$ is the chemical potential. The role of particle-hole asymmetry for thermoelectric transport is already visible in Drude-Boltzmann transport theory, where $\alpha=(2\pi^2/3) e T(\nu_\eps D_\eps)'$. In this expression, $\nu_\eps$ and $D_\eps$ denote the frequency-dependent density of states and diffusion coefficient, respectively, and the prime indicates a derivative with respect to frequency. In $2d$, and for a quadratic dispersion, the density of states and the disorder scattering time $\tau$ are approximately constant \cite{Fabrizio91,Schwiete21}. Correspondingly, the thermoelectric transport coefficient $\alpha$ is governed by the frequency-dependence of the diffusion coefficient, which in turn reflects the frequency dependence of the electron velocity, $D_\eps=v_\eps^2\tau/2$. The crucial role played by particle-hole asymmetry implies that a calculation of the thermopower has to be performed at a higher accuracy compared to electric or thermal conductivities. For the same reason, the most powerful analytical tools developed for the description of electronic transport in disordered systems, the conventional nonlinear sigma model (NL$\sigma$M) approach \cite{Wegner79,Efetov80,Finkelstein83} or the quasiclassical Green's function technique \cite{Larkin68}, are not straightforwardly applicable. We therefore make use of a recently derived generalized NL$\sigma$M for interacting systems, which includes particle-hole asymmetry \cite{Schwiete21}.

{\it Structure of the correlation function.---} Our calculation of the transport coefficients is based on the heat density-density correlation function $\chi_{kn}$, as in Ref.~\cite{Fabrizio91}. It is instructive to discuss the compatibility of the interaction corrections with the general structure of $\chi_{kn}$, which is strongly constraint by particle and energy conservation laws. The retarded heat density-density correlation function is defined as $\chi_{kn}(x_1,x_2)=-i\theta(t_1-t_2)\langle[\hat{k}(x_1),\hat{n}(x_2)]\rangle_T$, where $\hat{n}$ is the density operator, $\hat{k}=\hat{h}-\mu\hat{n}$ is the heat density operator, with Hamiltonian density $\hat{h}$ and chemical potential $\mu$, $x=(\bfr,t)$ combines spatial coordinates $\bfr$ and time $t$, and $\langle \dots\rangle_T$ denotes thermal averaging. As a function of momentum ${\bfq}$ and frequency $\omega$, the disorder averaged correlation function in the diffusive regime takes the following form \cite{Fabrizio91} 
\begin{align}
\chi_{kn}(\bfq,\omega)&=\frac{D_{n} \bfq^2 D_k \bfq^2 \chi^{st}_{kn}+i L\bfq^2 \omega}{(D_{n}\bfq^2-i\omega)(D_k\bfq^2-i\omega)}.\label{eq:chikngeneral}
\end{align}
In this equation, $D_{n}$ and $D_k$ are the diffusion coefficients for charge and heat, respectively, while $L$ is connected to the transport coefficient $\alpha$ and the Seebeck coefficient $S$ as $\alpha =eL/T$, $S=eL/(\sigma T)$. The static part of the correlation function is related to a thermodynamic susceptibility, $\chi^{st}_{kn}=-T\partial_T n$. The conservation laws for particle number and energy impose the constraint $\chi_{kn}(\bfq=0,\omega\rightarrow 0)=0$. For the $2d$ disordered electron gas with quadratic dispersion, and in the absence of interactions, the parameters are given as $D_n=D_k=D_0$, $\chi^{st}_{kn}=0$, and $L=Tc_0D_\eps'$, where $D_0=v_F^2\tau/2$, $D_\eps'=D_0/\mu$, and $c_0=2\pi^2 T\nu_0/3$ is the specific heat. In the presence of interaction corrections, the static part of the correlation function $\chi^{st}_{kn}$ depends on the parameter $z$ familiar from the renormalization of the Finkel'stein model \cite{Finkelstein83} as $\chi^{st}_{kn}=-c_0 T\partial_\mu z$ \cite{Fabrizio91,Schwiete21}. With this relation at hand, the leading order quantum corrections to the parameters $D_n$, $D_k$, and $L$ can be found from the dynamical part of the correlation function. 

The diffusion coefficients $D_{n}$ and $D_k$ are well known from studies of electric and thermal transport \cite{Finkelstein83,Castellani84,Castellani87,Schwiete14a, Schwiete14b}. The heat density-heat density correlation function takes the form $\chi_{kk}(\bfq,\omega)=-TcD_k\bfq^2/(D_k\bfq^2-i\omega)$, where $c$ is the specific heat, and allows to determine the thermal conductivity as $\kappa=cD_k$ \cite{Castellani87,Schwiete14a,Schwiete14b}. The charge diffusion coefficient, in turn, is known from the density-density correlation function $\chi_{nn}(\bfq,\omega)=-\partial_\mu n\, D_{n}\bfq^2/(D_{n}\bfq^2-i\omega)$, which is related to the electric conductivity as $\sigma=e^2 \partial_\mu n\, D_n$ \cite{Finkelstein83,Castellani84}. These relations provide tight constraints for the calculation of $\chi_{kn}$. Our result for $D_n=D_0+\delta D$, where $\delta D/D_0=-\rho \log 1/T\tau$, is fully consistent with the known RG result, $D_n=D/z_1$ \cite{Finkelstein83, Castellani84}, when applied to the Coulomb-only model, since $z_1=1$ in the absence of Fermi-liquid renormalizations. The case of the heat diffusion coefficient $D_k$ is more subtle. Here, we obtain \cite{d_RemarkSuppl} $D_k=(D_0+\delta D)/(1+\delta z)+I^h/2$, which is consistent with previous results for the diffusion of heat in the disordered electron liquid \cite{Schwiete16a,Schwiete16b}. The first term in the expression for $D_k$ reflects the relation $D_k=D/z$ obtained from the RG energy interval \cite{Castellani87, Schwiete14a,Schwiete14b}, with $\delta z=-\frac{1}{2}\rho\log{1}/{T\tau}$. The second term, which originates from subthermal energies, is responsible for the violation of the Wiedemann-Franz law in Eq.~\eqref{eq:dkappa} \cite{Schwiete16a,Schwiete16b}. The coefficient $L$ characterizing thermoelectric transport is obtained as $L=L_0\left(1-\frac{3}{4}I-I^h\right)$ with $I=\rho \log 1/T\tau$, which leads us to Eq.~\eqref{eq:alphaS}. We will discuss the different types of corrections to $L$ in more detail after introducing the formalism underlying our calculation. 

{\it NL$\sigma$M approach.---} The use of the NL$\sigma$M formalism for the calculation of $\chi_{kn}$ requires a generalization of the Finkel'stein model \cite{Finkelstein83} to include particle-hole asymmetry. It is further convenient to equip the model with potentials that can serve as source fields for generating the correlation function $\chi_{kn}$, namely a scalar potential coupling to the density, and Luttinger's gravitational potential \cite{Luttinger64} coupling to the heat density. The generalized Keldysh NL$\sigma$M for the disordered electron gas with particle-hole asymmetry in the presence of the scalar potential $\varphi$ and gravitational potential $\eta$ can be written as the sum of two terms \cite{Schwiete21}, $S=S_F[\hat{X}]+S_M[\hat{Q}]$, where 
\begin{align}
S_F[\hat{X}]=&\frac{i \pi\nu }{4}\tr\left[ D(\nabla \underline{\hat{X}})^2+4i \hat{\eps}^\eta_{\varphi} \underline{\delta \hat{X}} \right]\label{eq:SF}\\
&-\frac{\pi^2\nu^2}{4}\int_{\bfr,\bfr',\eps_i} \Big(\tr[\hat{\gamma}_i\{\hat{\lambda}(\bfr)\underline{\delta \hat{X}}(\bfr)\}_{\eps_1\eps_2}]\times\nonumber\\
&\times \hat{\gamma}_2^{ij}V_s(\bfr-\bfr') \tr[\hat{\gamma}_j \underline{\delta \hat{X}}_{\eps_3\eps_4}(\bfr')]\Big)\delta_{\eps_1-\eps_2,\eps_4-\eps_3},\nonumber\\
S_M[\hat{Q}]=&\frac{\pi\nu}{16}DD_\eps'\tr[\nabla^2 \hat{Q}(\nabla \hat{Q})^2].
\end{align}
The fields $\hat{X}$ and $\hat{Q}$ are matrices in Keldysh space carrying two spin and two frequency indices \cite{Kamenev11,Schwiete14,Schwiete21}. The field $\hat{Q}$ is familiar from the conventional NL$\sigma$M formalism without particle-hole asymmetry \cite{Wegner79,Efetov80,Finkelstein83} and fulfills the constraint $\hat{Q}^2=1$. It takes the form $\hat{Q}= \hat{U}\hat{\sigma}_3\hat{\bar{U}}$, where $\hat{U}\hat{\bar{U}}=1$ and $\hat{\sigma}_3$ denotes the third Pauli matrix in Keldysh space. The field $\hat{X}$ is related to $\hat{Q}$ as $\hat{X}=\hat{Q}+\frac{1}{4i}D_\eps'(\nabla{\hat{Q}})^2$. In Eq.~\eqref{eq:SF}, we also used the notation $\delta \hat{X}=\hat{X}-\hat{\sigma}_3$. The trace operation ``$\tr$" accounts for all degrees of freedom including the integration in ${\bf r}$, unless written explicitly. The frequency operator $\hat{\eps}_{\varphi}^\eta$ is defined as
$\hat{\eps}^\eta_{\varphi}=\frac{1}{2}\{\hat{\eps}-\hat{\varphi},\hat{\lambda}\}$, where $\hat{\lambda}=1/(1+\hat{\eta})$.
The matrix structure of the scalar fields is defined as in $\hat{\varphi}^l=\Sigma_{k=1,2}\varphi_k^l\hat{\gamma}_k$, where $\hat{\gamma}_1=\hat{\sigma}_0$ and $\hat{\gamma}_2=\hat{\sigma}_1$ are Pauli matrices in Keldysh space. The scalar fields also acquire a matrix structure in frequency space via $(\hat{\varphi}_{\bfr})_{\eps\eps'}=\hat{\varphi}_{\bfr,\eps-\eps'}$. Information about the occupation of states is encoded in the matrix 
\begin{align}
\hat{u}_\eps=\left(\ba{cc} 1&F_\eps\\ 0&-1\ea\right),\quad F_\eps=\tanh\frac{\eps}{2T},\label{eq:defu}
\end{align}
which enters the action through $\underline{\delta \hat{X}}=\hat{u}\delta \hat{X}\hat{u}$. The statically screened Coulomb interaction is denoted as $V_s$. In momentum space, it reads as $V_s(\bfq)=2\pi e^2/(|\bfq|+\kappa_s)$. 

The part of the action denoted as $S_F[\hat{X}]$ stands for the conventional Finkel'stein model \cite{Finkelstein83}, supplemented with source fields, and extended to include particle-hole asymmetry through the replacement $\hat{Q}\rightarrow \hat{X}$. The noninteracting part of $S_F$, in the absence of sources, takes the form $S_0[\hat{Q}]=\frac{i \pi\nu }{4}\tr[ D_{\hat{\eps}}(\nabla \hat{Q})^2+4i \hat{\eps} \hat{Q}]$. This action differs from the conventional model for noninteracting systems only by the frequency-dependence of the diffusion coefficient. Correspondingly, the frequency-dependent diffuson $\mathcal{D}^\eps_{\bfq,\omega}=(D_\eps \bfq^2-i\omega)^{-1}$, where $\eps$ is the center of mass frequency of the contributing retarded and advanced Green's functions, will form the basis of our perturbative calculation.  In a microscopic derivation, $S_F[\hat{X}]$ is found from a gradient expansion around the metallic saddle point $\hat{\sigma}_3$ without account of the massive modes and includes up to four gradients. By contrast, $S_M$ is obtained from a cumulant expansion including the coupling of soft and massive modes \cite{Wang94,Schwiete21}. It is worth noting that both $S_F$ and $S_M$ contain a term with four gradients, of the same form, albeit with different coefficients. In our calculation, the (combined) four-gradient term will contribute to Hikami-boxes, generalized to include particle-hole asymmetry. 

Due to the approximate constancy of the density of states, the terms representing particle-hole asymmetry in the NL$\sigma$M are all proportional to the derivative of the diffusion coefficient $D_\eps'$. These terms also differ from the terms present in the conventional model by their symmetry. The conventional Finkel'stein model in the absence of sources is invariant under the transformation $\hat{Q}\rightarrow \hat{Q}'$, where $\hat{Q}'_{\eps_1\eps_2}=-\sigma_2\hat{\sigma}_1\hat{Q}^t_{-\eps_1,-\eps_2}\hat{\sigma}_1\sigma_2$,
 $\hat{\sigma}_1$ is a Pauli matrix in Keldysh space, and $\sigma_2$ acts in spin space \cite{Schwiete21}. Each term in the action has a partner term in the generalized model, obtained via the replacement $\hat{Q}\rightarrow \hat{X}$ (with an additional change of coefficients for the four-gradient term due to the presence of $S_M$), which is proportional to $D_\eps'$ and odd under the transformation $\hat{Q}\rightarrow \hat{Q}'$. This has important consequences. Extending the symmetry analysis to the source terms \cite{Schwiete21}, one finds that a nonvanishing result for $\chi_{kn}$ can only be obtained by including terms with $D_\eps'$ in the analysis.

{\it Calculation.---} The correlation functions $\chi_{kn}$ in the diffusive limit can be obtained from the Keldysh partition function $\mathcal{Z}=\int DQ \exp(iS)$ as the functional derivative $\chi_{kn}(x_1,x_2)=(i/2)\delta^2 \mathcal{Z}/\delta\eta_2(x_1)\delta \varphi_1(x_2)$ evaluated at $\vec{\eta}=\vec{\varphi}=0$. The particle-hole asymmetry is explicit through the $D_\eps'$-dependence of the action. The generalized NLSM with source fields $\hat{\eta}$ and $\hat{\varphi}$ therefore allows us to formulate a systematic linear response theory for the thermoelectric transport coefficient. The static part of the correlation function has already been analyzed in Refs.~\cite{Fabrizio91} and \cite{Schwiete21}. Here, we will focus on the dynamical part.

\begin{figure}[tb]
\includegraphics[width=8.5cm]{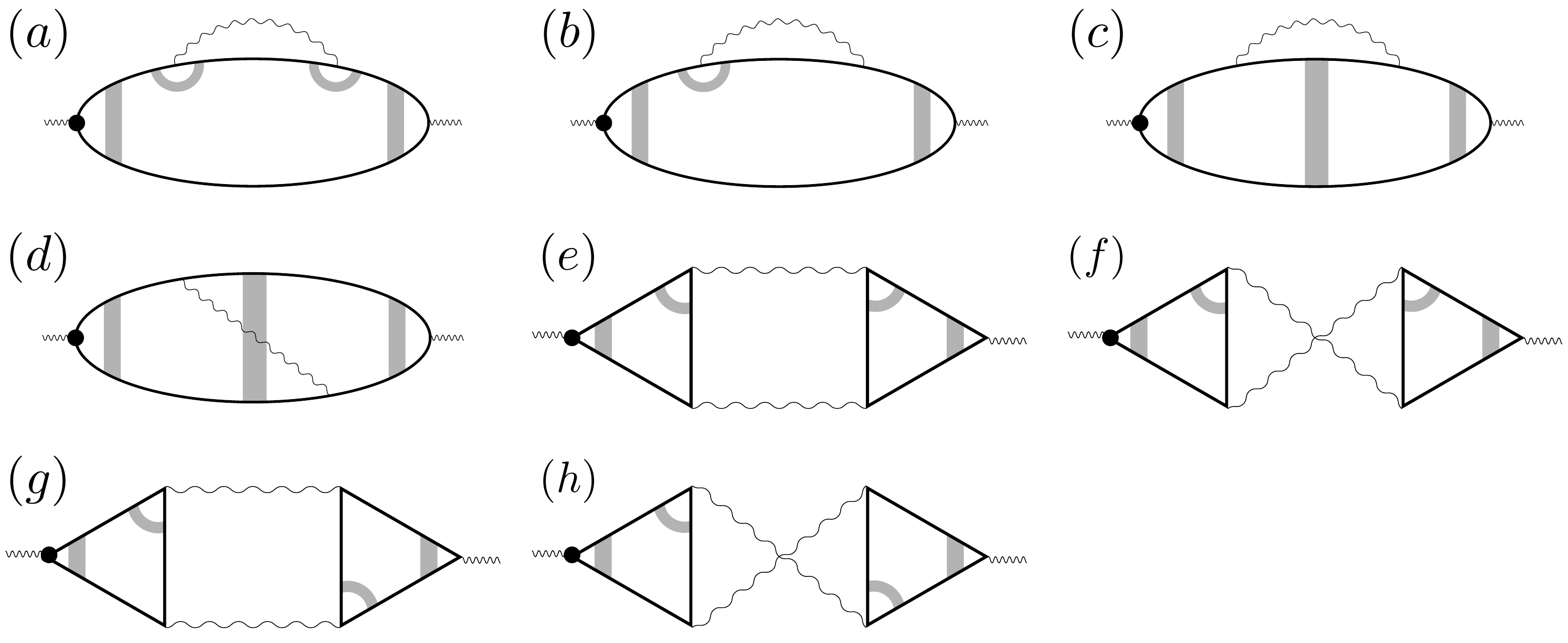}
\caption{Interaction corrections to the diffuson \cite{f_RemarkDiagram1}. Solid lines represent electronic Green's functions, wiggly lines the dynamically screened Coulomb interaction, and shaded rectangles or arcs the diffuson $\mathcal{D}_{\eps}(\bfq,\omega)$. The full circle indicates heat density vertices and is shown on the left, the density vertex is on the right. Each diagram represents a class of two (in the case of $(b)$ four) symmetry-related diagrams.}
\label{fig:diagrams_diffusion}
\end{figure}

A first order expansion of $\hat{\eps}_{\varphi}^\eta$ in $\eta_2$ and $\varphi_1$ is sufficient for finding the dynamical part of the heat density-density correlation function, $\chi^{dyn}_{kn}=\chi_{kn}-\chi_{kn}^{stat}$. For the perturbative calculation, we employ a one-parameter family of parameterizations of the $\hat{Q}$ matrix with unit Jacobian, $\hat{Q}=\hat{\sigma}_3f(\hat{P})$, where $f(x)=(x/2+\sqrt{1+(1-\lambda)x^2/4})^2/(1-\lambda x^2/4)$ \cite{Ivanov06}. The independence of the results for the correlation function on the parameter $\lambda$ serves as useful consistency check \cite{e_RemarkParameterization}. Relevant diagrams for the corrections to the diffusion propagator are displayed in Fig.~\ref{fig:diagrams_diffusion}. 
While heat vertex corrections and charge vertex corrections are crucial for establishing the structure of the correlation function $\chi_{kn}$ in Eq.~\eqref{eq:chikngeneral}, the corrections to $L$ (and thereby to $\alpha$ and $S$) are found from the diagrams displayed in Fig.~\eqref{fig:diagrams_diffusion}. For each diagram, an expansion in $D'_\eps$ is required in order to obtain a finite result. This expansion may originate from different terms, namely from diffusons $\mathcal{D}^\eps_{\bfq,\omega}$, from the generalized Hikami box terms through $\Tr[D_\eps \nabla^2 \hat{Q}]$ or $DD'_\eps\Tr[\nabla^2 \hat{Q}(\nabla \hat{Q})^2]$, or from source and interaction terms that become $D_\eps'$-dependent via the replacement $\hat{Q}\rightarrow \hat{X}$. The advantage of the described method is that the particle-hole asymmetry is explicit, so that a systematic expansion can be achieved straightforwardly. On the other hand, each diagram gives rise to several distinct contributions, which need to be added consistently. Fortunately, the structure of the correlation function imposes tight constraints on the calculation. In particular, the results for $D_n$ and $D_k$ obtained from the calculation of $\chi_{kn}$ can be checked against those known from studies of $\chi_{nn}$ and $\chi_{kk}$ as discussed above. 

The result for $\delta \alpha$ can conveniently be expressed through the corrections to the frequency-dependent diffusion coefficient \cite{d_RemarkSuppl},
\begin{align}
\delta \alpha = ec_0\delta D_\eps',\quad \delta D_\eps=\delta D^{(1)}_\eps+\delta D^{(2)}_\eps+\delta D_\eps^h.\label{eq:dalpha}
\end{align} 
with $\eps \lesssim T$ assumed. The first two terms in this sum are contributions from the RG interval of energies. In particular, $\delta D^{(1)}_\eps$, is a generalization of the conventional interaction correction $\delta D$ to the diffusion coefficient related to Fig.~\ref{fig:diagrams_diffusion}(c) \cite{g_RemarkIntegrals}
\begin{align}
\delta D=iD\int_{\bfk,\nu} \Delta_{\eps,\nu}V^R_{\bfk,\nu} D\bfk^2\mathcal{D}_{\bfk,\nu}^3.\label{S12}
\end{align}
In this formula, the combination of distribution functions $\Delta_{\eps,\nu}=\mathcal{F}_{\eps+\nu}-\mathcal{F}_{\eps-\nu}$ favors large frequencies $\nu> T$. The correction $\delta D^{(1)}_\eps$ includes $\delta D$, but also incorporates the frequency-dependence of the bare diffusion coefficient as 
\begin{align}
\delta D_\eps^{(1)}=iD_\eps\int\Delta_{\eps,\nu} V^R_{\bfk,\nu}D_\eps\bfk^2 (\mathcal{D}_{\bfk,\nu}^\eps)^3\label{eq:dD1}.
\end{align}
The correction $\delta D$ is universal in the sense that it does not depend on $D$. Correspondingly, if we replace $V^R_{\bfk,\nu}$ by $V^R_{\bfk,\nu;\eps}=(V_s(\bfk)+\Pi_{\bfk,\nu;\eps})^{-1}$ with $\Pi_{\bfk,\nu;\eps}\equiv 2\nu_0 D_\eps \bfk^2 \mathcal{D}^\eps_{\bfk,\nu}$ in the equation for $\delta D_\eps^{(1)}$, so that all diffusion coefficients on the right hand side depend on the same frequency $\eps$, then $\delta D_\eps^{(1)}$ coincides with $\delta D$. We can therefore write 
\begin{align}
\delta D^{(1)}_\eps=\delta D-iD_\eps\int_{\bfk,\nu}\Delta_{\eps,\nu} \delta V^R_{\bfk,\nu;\eps}D_\eps\bfk^2 (\mathcal{D}_{\bfk,\nu}^\eps)^3\label{S14}
\end{align}
with $\delta V^R_{\bfk,\nu;\eps}=V^R_{\bfk,\nu;\eps}-V^R_{\bfk,\nu}$. Since $\delta V^R$ is already of first order in $\eps$, we may neglect the $\eps$ dependence of all other terms in the expression, expand to first order in $\eps$, and evaluate the resulting logarithmic integral (in the universal limit $V^R\approx \Pi^{-1}$)
\begin{align}
\delta D^{(1)}_\eps-\delta D=\eps D'_\eps \int_{\bfk,\nu} \Delta_{\eps,\nu} \nu V^R_{\bfk,\nu} D\bfk^2 \mathcal{D}_{\bfk,\nu}^4=
-\frac{1}{2}\eps D_\eps'I. \label{S15}
\end{align}
This correction was absent in Ref.~\cite{Fabrizio91}. 

The second term in the expression for $\delta D_\varepsilon$ in Eq.~\eqref{eq:dalpha}, $\delta D^{(2)}_\eps$, is obtained from the generalized Hikami-box diagrams Fig.~\ref{fig:diagrams_diffusion}$(a)$ and Fig.~\ref{fig:diagrams_diffusion}$(b)$. The contribution of Fig.~1$(a)$ has its origin in the total four-gradient term in the action, $\delta S^{tot}_{4}=-S_M$. It takes the form 
\begin{align}
\delta  D^{(2a)}_\eps=\frac{1}{4}D_\eps '\int_{\bfk,\nu} \bar{\Delta}_{\eps,\nu} V^R_{\bfk,\nu}D\bfk^2 \mathcal{D}_{\bfk,\nu}^2
\end{align}
with $\bar{\Delta}_{\eps\nu}=\mathcal{F}_{\eps+\nu}+\mathcal{F}_{\eps-\nu}$. The contribution of the diagram in Fig.~1$(b)$ arises through the $D_\eps'$-dependence of $\delta \hat{X}$ in the interaction part of $S_F[\hat{X}]$, $\delta D_\eps^{(2b)}=-\frac{1}{2} D'_\eps\int_{\bfk,\nu} \bar{\Delta}_{\eps,\nu} V^R_{\bfk,\nu} D\bfk^2 \mathcal{D}_{\bfk,\nu}^2=-2\delta D^{(2a)} $. We confirmed that in a conventional diagrammatic calculation the total correction 
\begin{align}
\delta D^{(2)}_\eps=\delta D_\eps^{(2a)}+\delta  D^{(2b)}_\eps=-\frac{1}{4}D'_\eps \eps I\label{S17}
\end{align} is entirely obtained from a careful expansion of the Hikami-box diagram with account of the particle-hole asymmetry \cite{d_RemarkSuppl}. This correction has also been identified in Ref.~\cite{Fabrizio91}, with the same result. 

The term $\delta D_\eps^h$ in Eq.~\eqref{eq:dalpha} represents the interaction corrections from the sub-thermal energy interval. All the diagrams displayed in Fig.~\ref{fig:diagrams_diffusion} contribute to these corrections \cite{d_RemarkSuppl}. Here, we discuss one of the contributions, $\delta D_\eps^{h(a)}$, which is obtained from diagrams \ref{fig:diagrams_diffusion}$(e)$-\ref{fig:diagrams_diffusion}$(h)$ 
\begin{align}
\delta D_\eps^{h(a)}
=-i D_\eps'\int_{\bfk,\nu}\bar{\Delta}_{\eps,\nu}\Im V^R_{\bfk,\nu} \mathcal{D}_{\bfk,\nu}
=- D_\eps'\eps I^h.
\end{align}
It is worth elaborating on the specific form of the integrand. Under the integral, the function $\bar{\Delta}_{\eps,\nu}=\mathcal{F}_{\eps+\nu}+\mathcal{F}_{\eps-\nu}$ restricts the range of relevant frequencies  to $|\nu|\lesssim T$. For momenta $k$ fulfilling the inequalities $|\nu|/D\kappa_s<k<\sqrt{|\nu|/D}$, the imaginary part of the interaction can be approximated as $\Im V^R_{\bfk,\nu}\approx- \frac{1}{2\nu_0}\frac{\nu}{D\bfk^2}$. It is this bare $1/D\bfk^2$ singularity that gives rise to the logarithmic correction. The same range of momenta is also responsible for the double-logarithmic temperature-dependence of the tunneling density of states \cite{AltshulerLee80,AltLee80} as well as spurious double-logarithmic contributions that appear at intermediate steps of the calculation for the correlation function. In contrast to these examples, both $|\nu|$ and $D\bfk^2$ in the expression for $\delta D^{h(a)}_\eps$ are bound to be smaller than $T$. This is why only a single logarithm arises, and the contribution falls outside of the RG range of energies. 

The total correction from the subthermal energy interval coincides with $\delta D_{\eps}^{h(a)}$, namely $\delta D_\eps^h=-D'_\eps \eps I^h$ \cite{d_RemarkSuppl}. After combining this result with $\delta D_\eps^{(1)}$ and $\delta D_\eps^{(2)}$, Eq.~\eqref{eq:dalpha} leads directly to 
\begin{align}
\quad \frac{\delta \alpha}{\alpha}=\frac{3}{4}\frac{\delta \sigma}{\sigma}-I^h.
\end{align}
We see that both types of logarithmic corrections contribute to a decrease of $\delta \alpha/\alpha$. Our main result for the interaction correction to the thermopower $\delta S/S$, stated in Eq.~\eqref{eq:alphaS}, follows immediately.

{\it Discussion.---} (i) As argued in Ref.~\cite{Castellani88a}, weak localization effects do not affect $\alpha$ at first order in the dimensionless resistance $\rho$. As long as quantum corrections remain small, the temperature-dependence of $\alpha/T$ can therefore be expected to be dominated by the interaction corrections discussed in this manuscript. Weak localization corrections do contribute to the thermopower $S=\alpha/\sigma$ through $\delta \sigma$. These corrections have been studied via their magnetic field dependence in Ref.~\cite{Rafael04}.\\
(ii) Thermopower measurements are well within experimental capabilities \cite{Fletcher01,Rafael04,Mokashi12}. Refs.~\cite{Fletcher01} and \cite{Mokashi12} primarily addressed strong correlation effects in the vicinity of the metal-insulator transition. Finding the thermopower on the metallic side of the metal-insulator transition theoretically will require an extension of the presented approach to a full RG analysis, combined with a calculation of subthermal corrections with the renormalized action, a program previously implemented for the thermal conductivity \cite{Schwiete14b,Schwiete16a,Schwiete16b}.

{\it Acknowledgments.---} The authors would like to thank A.~M.~Finkel'stein and T.~Micklitz for discussions, and K.~Michaeli for collaboration during the early stages of this work. This work was supported by the National Science Foundation (NSF) under Grant No. DMR-1742752.


\begin{thebibliography}{52}
\bibitem [{a_R()}]{a_RemarkThermal}%
  \BibitemOpen
  \href@noop {} {}\bibinfo {note} {Here, we do not distinguish $\kappa$ defined
  for ${\bf j}=0$ and for ${\bf E}=0$ (as in this matrix), because the
  difference is negligibly small for typical metallic systems at low
  $T$.}\BibitemShut {Stop}%
\bibitem [{b_R()}]{b_RemarkAmplitudes}%
  \BibitemOpen
  \href@noop {} {}\bibinfo {note} {We assume that Fermi liquid interaction
  amplitudes are small.}\BibitemShut {Stop}%
\bibitem [{c_R()}]{c_RemarkHsuFabrizio1}%
  \BibitemOpen
  \href@noop {} {}\bibinfo {note} {Ref.~\onlinecite{Hsu89} reported a
  temperature-independent leading correction to $\alpha$, in contrast to
  Ref.~\onlinecite{Fabrizio91} and our work. Since Ref.~\onlinecite{Hsu89}
  determined $\alpha$ from the heat current-current correlation function, a
  direct comparison of the calculations is not possible. However,
  Ref.~\onlinecite{Fabrizio91} suggested that specific terms were missing in
  Ref.~\onlinecite{Hsu89}.}\BibitemShut {Stop}%
\bibitem [{e_R()}]{e_RemarkParameterization}%
  \BibitemOpen
  \href@noop {} {}\bibinfo {note} {Up to order $\hat{P}^4$ in the expansion of
  $\hat{Q}$, which is sufficient for our calculation, the case $\lambda=-1$
  coincides with the widely used square-root even parametrization
  $f(x)=\sqrt{1+x^2}+x$, which corresponds to the diagrammatic cross-technique
  for the disorder averaging.}\BibitemShut {Stop}%
\bibitem [{f_R()}]{f_RemarkDiagram1}%
  \BibitemOpen
  \href@noop {} {}\bibinfo {note} {These diagrams visualize the NL$\sigma$M
  calculation. For the convenience of the reader, they are drawn in a form
  familiar from conventional diagrammatic perturbation theory
  \cite{e_RemarkParameterization, d_RemarkSuppl}}\BibitemShut {NoStop}%
\bibitem [{g_R()}]{g_RemarkIntegrals}%
  \BibitemOpen
  \href@noop {} {}\bibinfo {note} {We use the notation $\int_\nu=\frac{1}{2\pi}
  \int d\nu$ and $\int_{\bfk}=\frac{1}{(2\pi)^2} \int d^2 k$.}\BibitemShut
  {Stop}%
\end{thebibliography}

%

\end{document}